# Negative Magnetoresistance in Dirac Semimetal Cd$_3$As$_2$

Hui Li[1]*, Hongtao He[2]*, Hai-Zhou Lu[2]*, Huachen Zhang[1], Hongchao Liu[1], Rong Ma[1], Zhiyong Fan[3], Shun-Qing Shen[4]**, Jiannong Wang[1]**

*[1]Department of Physics, the Hong Kong University of Science and Technology, Clear Water Bay, Hong Kong, China*

*[2]Department of Physics, South University of Science and Technology of China, Shenzhen, Guangdong 518055, China*

*[3]Department of Electronics and Computer Engineering, the Hong Kong University of Science and Technology, Clear Water Bay, Hong Kong, China*

*[4]Department of Physics, the University of Hong Kong, Pokfulam Road, Hong Kong, China*

* These authors contributed equally to this work.

** Correspondence and requests for materials should be addressed to S.S. (email: sshen@hku.hk) and J. W. (email: phjwang@ust.hk).

**Abstract.** A large negative magnetoresistance is anticipated in topological semimetals in the parallel magnetic and electric field configuration as a consequence of the nontrivial topological properties. The negative magnetoresistance is believed to demonstrate the chiral anomaly, a long-sought high-energy physics effect, in solid-state systems. Recent experiments reveal that Cd$_3$As$_2$, a Dirac topological semimetal, has the record-high mobility and exhibits positive linear magnetoresistance in the orthogonal magnetic and electric field configuration. However, the negative magnetoresistance in the parallel magnetic and electric field configuration remains unveiled. Here, we report the observation of the negative magnetoresistance in Cd$_3$As$_2$ microribbons in the parallel magnetic and electric field configuration as large as 66% at 50 K and even visible at room temperatures. The observed negative magnetoresistance is sensitive to the angle between magnetic and electrical field, robust against temperature, and dependent on the carrier density. We have found that carrier densities of our Cd$_3$As$_2$ samples obey an Arrhenius's law, decreasing from $3.0\times10^{17}$ cm$^{-3}$ at 300 K to $2.2\times10^{16}$ cm$^{-3}$ below 50 K. The low carrier densities result in the large values of the negative magnetoresistance. We therefore attribute the observed negative magnetoresistance to the chiral anomaly. Furthermore, in the perpendicular magnetic and electric field configuration a positive non-saturating linear magnetoresistance up to 1670% at 14 T and 2 K is also observed. This work demonstrates potential applications of topological semimetals in magnetic devices.

Searching for the signature of the Adler-Bell-Jackiw chiral anomaly[1-3] in three-dimensional topological semimetals is one of the focuses in condensed matter physics. The topological semimetals have a band structure with the conduction and valence energy bands touching at a finite number of paired Weyl nodes[4-6] (Fig. 1a). In each pair, the two Weyl nodes carry opposite chirality, and paired monopoles and anti-monopoles of Berry curvature in momentum space[7] (Fig. 1b). The nontrivial Berry curvature can couple an external magnetic field to the velocity of electrons, leading to a chiral current that is linearly proportional to the field. The correlation of chiral currents further contributes to an extra conductivity that quadratically grows with increasing magnetic field, in a magnetic field and an electric field applied parallel to each other. This $B^2$ positive conductivity in weak parallel magnetic fields, or negative magnetoresistance (negative MR), is rare in non-ferromagnetic materials, thus can serve as one of the transport signatures of the topological semimetals. More importantly, because of its relation to the chiral charge pumping between paired Weyl nodes, the negative magnetoresistance is also believed to be a signature of the chiral anomaly[8,9].

Among the recently identified candidates for topological semimetals (e.g., $(Bi_{1-x}In_x)_2Se_3$[10], $Na_3Bi$[11-13], $TlBiSSe$[14], $TaAs$[15-18]), the Dirac semimetal $Cd_3As_2$[19-24] has peculiar transport properties, such as a giant MR in perpendicular magnetic fields and record high mobility[25-30], thus is of great potential in device applications. The negative MR possibly associated with the chiral anomaly has been claimed in several topological semimetals, including BiSb alloy[31], $ZrTe_5$[32], $TaAs$[33,34], and $Na_3Bi$[35]. However, the chiral anomaly in $Cd_3As_2$ is not yet observed. One of the reasons is that the carrier density in earlier samples was too high (over $10^{18}/cm^3$). The chiral anomaly arises because of the nontrivial Berry curvature, which diverges at the Weyl nodes, so the Fermi energy $E_F$ has to be as close to the Weyl nodes as possible for a clear signal of the negative MR.

In this work, we systematically investigate the magnetotransport properties of $Cd_3As_2$ microribbons, in which the carrier density is found to obey an Arrhenius's law, decreasing from $3.0\times10^{17}$ cm$^{-3}$ at 300 K to $2.2\times10^{16}$ cm$^{-3}$ below 50 K. In perpendicular magnetic fields, the ribbon exhibits a very large non-saturating positive linear MR (linear MR) up to 300 K. In contrast, when the magnetic field is applied in parallel with the measurement electric field, a negative MR is observed. It is sensitive to the angle between magnetic and electrical field and shows a parabolic dependence on the low magnetic fields (< 1 T) and persists up to 300 K. More importantly, our analysis reveals a characteristic carrier density dependence of the observed negative MR which is in agreement with the semiclassical theory about the chiral anomaly in topological semimetals. All the experimental evidence makes us believe that the chiral anomaly induced negative MR is indeed realized in our $Cd_3As_2$ ribbons with low carrier density. By studying the carrier density dependence of the observed linear MR in perpendicular magnetic fields, possible physical origins are also discussed. Our work shows that the carrier density plays an important role in the observation of the negative and linear MR and their magnitudes.

## Results

**Device characteristics.** Fig. 2 (a) shows the scanning electron microscopy (SEM) image of the four-terminal $Cd_3As_2$ devices studied in this work. The width $w$ and inter-voltage-probe distance $l$ are 1210 and 1600 nm, respectively. According to the atomic force microscopy measurement shown in Fig. 2(b), the ribbon thickness $t$ is about 327 nm. Figure 2(c) shows the measured temperature ($T$) dependence of the resistance ($R$) of the $Cd_3As_2$ ribbon. With decreasing temperature, the ribbon changes from an insulating behavior to a metallic one, with a resistance peak appearing around 50 K. We note that similar $R$-$T$ curves were also observed in recent studies of Dirac semimetals, where chiral anomaly induced transport features were reported[36,37]. One of the important properties of Dirac semimetals is the giant non-saturating linear magnetoresistance (linear MR) in high magnetic fields[25-30]. Indeed, our $Cd_3As_2$ ribbons do exhibit such an intriguing linear MR. Fig. 2 (d) shows the MR measured at $T$ = 2 K with varying angle between magnetic and electric field applied (see inset). When the magnetic field ($B$) is applied perpendicular to the ribbon in the $z$ direction, $i.e.$ the $B$-field tilting angle $\theta = 90^o$, a linear MR up to 1670% at 14 T is observed. This linear MR decreases at smaller titling angles when we rotate the $B$-field in the $x$-$z$ plane. For $\theta \leq 10^o$, a negative MR begins to emerge in low magnetic fields, as shown in the Fig. 2 (e). For $\theta = 0^o$, $i.e.$ the $B$-field is parallel to the electric field direction, the negative MR is the most prominent. The emergence of such a negative MR when $B//E$ is recently ascribed to the chiral anomaly of topological semimetals[8,9].

**Linear MR in perpendicular fields.** To gain more physical insights into the observed linear MR and negative MR, we further study the temperature dependence of these two phenomena. Fig. 3 (a) shows the *R-B* curves measured at $\theta = 90^o$ and different temperatures indicated. As temperature increases from 2 K to 50 K, the linear MR in high magnetic fields (> 8 T) is almost unchanged but it starts to weaken gradually as *T* further increases. At low temperatures, some wiggles are superimposed on the linear MR but quickly disappear as temperature increases. In addition, we note that the *R-B* curves show a quadratic *B*-field dependence in low magnetic fields (< 1 T) distinct from the linear MR in high magnetic fields. As shown in the inset of Fig. 3 (b), the *R-B* curves at *T* = 2 K to 300 K can be well fitted by parabolas within a small *B*-field range. Such a parabolic MR is believed to arise from the Lorentz deflection of carriers and the $B^2$ fitting allows us to deduce the carrier mobility, $\mu$, by[38]

$$R(B) = R_0[1 + (\mu B)^2]. \qquad (1)$$

The obtained mobility of our $Cd_3As_2$ ribbon at different temperatures is shown in Fig. 3(b). It increases monotonically with decreasing temperature. At *T* = 2 K, the mobility reaches $10^4$ cm$^2$V$^{-1}$s$^{-1}$, which is comparable to those reported in previous studies of $Cd_3As_2$[26,27,29,30]. Based on the Drude model of electric conduction, the carrier density *n* of our ribbon can be derived by

$$n = 1/(\rho e \mu) = l/(Rwte\mu), \qquad (2)$$

where *e* is the elementary charge, $\rho$ is the resistivity of the ribbon, and the resistance *R* at different temperatures is shown in Fig. 2 (c). Figure 3 (c) shows the obtained carrier density as a function of temperature. Below 50 K, the carrier density is almost a constant. From 50 K to 300 K, it increases with temperature by about one order of magnitude from $2.2\times10^{16}$ cm$^{-3}$ to $3.0\times10^{17}$ cm$^{-3}$, following Arrhenius's law [solid curve in Fig. 3 (c)]:

$$n(T) \propto \exp(-\Delta/k_B T), \qquad (3)$$

where $k_B$ is the Boltzmann constant and the thermal activation energy $\Delta$ is about 51 meV. Such a thermally activated process of carriers accounts for the insulator-like *R-T* behavior above 50 K shown in Fig. 2 (c) while the metallic behavior below 50 K is mainly due to the increase of the carrier mobility with decreasing temperature. Since the intrinsic carriers will not follow Arrhenius's law for Dirac fermions[39], we tentatively ascribe the thermal activation of carriers to some traps present in our $Cd_3As_2$ ribbons. It is worth pointing out that both the observed linear MR and the carrier density are almost temperature independent when $T < 50$ K and become temperature dependent when $T > 50$ K. This coincidence implies that the observed strength of the linear MR is related to the change of carrier density. We can thus examine the temperature dependence of the linear MR in the temperature range from 50 K to 300 K in terms of the carrier density. The slope $k$ of the linear MR is extracted by linearly fitting the $R$-$B$ curves between 8 and 14 T, as shown in Fig. 3 (a). The obtained $k$ as a function of $n$ is shown in Fig. 3 (d).

A giant non-saturating linear MR has been observed in various Dirac semimetals[25-30], but its physical origin is still under debate. One possible mechanism is the quantum linear MR model proposed by Abrisokov, where a non-saturating linear MR would appear in three-dimensional gapless semiconductors with linear energy dispersion when all the carriers are condensed into the lowest bands of Landau levels, *i.e.* in the quantum limit[40]. According to this model, the linear MR is temperature independent but should follow the $1/n^2$ dependence, *i.e.* $\rho(B_\perp) \propto B_\perp/n^2$, where $n^2$ arises because the longitudinal resistance $\rho \approx \sigma/\sigma_H{}^2$ in the limit that the longitudinal conductivity $\sigma$ is much smaller than the Hall conductivity $\sigma_H$, and the Hall conductivity is proportional to $n$. However, considering that the longitudinal $\sigma$ is also proportional to $n$, the carrier density dependence should be corrected to $\rho(B_\perp) \propto B_\perp/n$. As a result, the slope of the quantum linear MR is inversely proportional to the carrier density. For comparison we plot $k \propto 1/n$ (red solid line) & $k \propto 1/n^2$ (blue solid line) curves in Fig. 3 (d). As it can be seen, the measured $k$ follows $1/n$ dependence. In addition, our system is believed to enter the quantum limit in the fitting magnetic field range from 8 to 14 T, as will be discussed later in this work. All these seem to suggest the quantum model as the underlying physical origin of the observed linear MR in high fields shown in Fig. 3 (a). Besides this quantum model, there is another classical model proposed by Parish and Littlewood to account for the linear MR observed in polycrystalline silver chalcogenides[41]. It is the disorder-induced admixture of the Hall signal that gives rise to the linear MR.

Considering the low carrier density and quite high mobility of our $Cd_3As_2$ ribbons, this classical model is unlikely applicable to the observed linear MR.

**Negative MR in parallel fields.** In Fig. 4 (a), the $R$-$B$ curves obtained at $\theta = 0^o$ are shown at different temperatures. There exists a critical $B$ field for each curve, below which a pronounced negative MR is observed even with temperature up to 300 K. At $T =$ 50 K and $B = 8$ T, the highest negative MR of 66% can be obtained. It is also noted in Fig. 4 (a) that some MR ripples are superimposed on the negative MR especially as low temperatures. Besides the apparent negative MR, a small resistance dip appears in low fields with $T < 20$ K, as shown in Fig. 4 (b). Such a dip is believed to arise from the weak anti-localization effect in Dirac semimetals[31,42]. In Fig. 4 (c), the low-field negative MR is shown at different temperatures. Note that the measured resistance has been converted to the conductivity $\sigma$ and $\Delta\sigma = \sigma - \sigma_0$, where $\sigma_0$ is the conductivity at zero magnetic field. Remarkably, all the data in Fig. 4 (c) shows a quadratic dependence on the magnetic field, as indicated by the fitting curves with $\Delta\sigma = C_a B^2$. This is consistent with the prediction in previous theoretical studies[8,9]. When an external electric field is applied in parallel with the magnetic field, chiral charges will be pumped from one Weyl node to the other, as a result of the nontrivial Berry curvature [see Fig. 1(b)]. The chiral charge is therefore not conserved at each Weyl node. Such a chiral anomaly is expected to give rise to a prominent positive conductivity proportional to $B^2$ based on a semiclassical transport calculation[8,9]. It should be noted that such a quadratic field dependence of the conductivity is only valid in weak magnetic fields or high temperatures such that the Landau quantization can be ignored. Indeed, the quadratic field dependence of $\Delta\sigma$ in Fig. 4(c) occurs in low magnetic fields ($< 0.8$ T).

Furthermore, because the chiral anomaly arises from the nontrivial Berry curvature, which diverges at the Weyl nodes, the positive conductivity will increase with decreasing Fermi energy and carrier density. More specifically, the theory predicts $C_a \propto E_F^{-2} \propto n^{-2/3}$ if the Fermi level $E_F$ is close to the Weyl nodes[8,9]. As a result, we plot the obtained fitting parameter $C_a$ as a function of the carrier density in Fig. 4 (d) together with a $C_a \propto n^{-2/3}$ curve (solid red line). One can see that $C_a$ does obey the $n^{-2/3}$ relationship reasonably well. All these experimental evidences, i.e. measured $\Delta\sigma = C_a B^2$ and $C_a \propto n^{-2/3}$, lead us to believe that it is the chiral anomaly that gives rise to the observed negative MR shown in Fig. 4(a). In the above discussion, we only consider the influence of $n$ on $C_a$. But $C_a$ is also proportional to the inter-node scattering time $\tau$ [8,9]. Since the momentum transfer assisted by phonons is suppressed at low temperatures, $\tau$ will increase with decreasing temperature. As shown in Fig. 3 (c), the carrier density of our sample follows Arrhenius's law. Different $\tau$ would be expected for different carrier density $n$, or $\tau$ increases with decreasing $n$. Therefore, the $n^{-2/3}$ fitting of $C_a$ in Fig. 4 (d), which assumes a constant $\tau$ for different $n$, would over- and underestimate the value of $C_a$ at high and low carrier densities, respectively. This accounts for the deviation of $C_a$ from the fitting at 300 K and below 50 K shown in Fig. 4 (d).

It is well known that, in solids, negative MR may have other physical origins. It can arise from the weak localization effect due to the quantum interference of time-reversed scattering loops. Since the phase coherence length decreases rapidly with increasing temperature, the weak localization can only be observed at low temperatures. The negative MR observed in $Cd_3As_2$ ribbons persists up to 300 K, therefore cannot be attributed to the weak localization effect. Magnetic scattering might be another mechanism for negative MR, as reported in some magnetic systems[43], but our $Cd_3As_2$ ribbons are non-magnetic.

**MR in the quantum limit.** As mentioned above, a critical magnetic field $B_C$ exists for each MR curve to characterize the change of MR from negative to positive in Fig. 4 (a) as illustrated by the arrows. We believe that this sign change in MR is an indication of the system to enter the quantum limit at $B_C$, *i.e.*, the Fermi energy crosses only the lowest Landau bands with the band index $\nu = 0$ and lies right below the bottom of the $\nu = 1$ Landau bands, as shown in the inset of Fig 5(a). The energy spacing between the $\nu = 1$ and $\nu = 0$ bands is roughly related to the cyclotron frequency as $\hbar\omega = \hbar eB/m^*$, with $m^*$ as the effective mass of carriers in $Cd_3As_2$. On the other hand, $\hbar\omega = \hbar^2 k_F^2/(2m^*)$, where the Fermi wave vector in the $\nu = 0$ bands is related to the carrier density according to $k_F = 2\pi^2\hbar n/(eB)$. Combing them, we can deduce a relationship between the critical field and the carrier density as $B_C = (2\pi^4)^{1/3}\hbar n^{2/3}/e \approx 3.8 * 10^{-11} n^{2/3}$. In order to justify this scenario, we extract $B_C$ from the MR curves at different temperatures (except 300 K curve) in Fig. 4 (a) and plot them in terms of the carrier density $n^{2/3}$ as it is shown in Fig. 5 (a). As the carrier density only changes within the temperature range of 50 K – 200 K, a straight line fitting of the data in this range with $B_C = \beta n^{2/3}$ yields $\beta = 3.0 * 10^{-11}$, which is close to the above theoretical value. Such a good agreement strongly supports our assumption that our system is indeed in the quantum limit above $B_C$. The deviation of data below 50 K, where the carrier density is almost constant, is probably caused by the superimposed ripples on the MR which prevent an accurate extraction of $B_C$. As it is shown in Fig. 4 (a), after the system is in the quantum limit, the measured magnetoresistance becomes positive, or the magnetoconductivity becomes negative. We find at high fields approaching 14 T the magnetoconductivity follows a good $B^{-1}$ dependence, as indicated by the linear fittings in Fig. 5 (b) red-solid curves. We also find that the obtained slope $k'$ of these linear fittings is inversely proportional to the carrier density $n$ [see Fig. 5 (c)]. This observed negative linear magnetoconductivity at high fields is contrary to the theoretical expectation, in which a positive linear magnetoconductivity is predicted as an additional signature of the chiral anomaly[8,44,45]. However, in reality the relaxation time and Fermi velocity can bring extra magnetic field dependences, leading to either positive or negative magnetoconductivity in Dirac semimetals[46].

## Discussion

In conclusion, we have performed a systematic magnetotransport study of $Cd_3As_2$ microribbons. Due to the low carrier density in our samples, we can observe both the non-saturating linear MR in a high perpendicular magnetic field and the quadratic negative MR when a weak magnetic field is in parallel with the measurement electric field. Furthermore, the thermally activated behavior of carriers in our $Cd_3As_2$ ribbons allows us to study the carrier density dependence of these two phenomena. Although the mechanism for the linear MR is still not well understood, the quadratic negative MR can be safely ascribed to the chiral anomaly intrinsic to the Dirac semimetal $Cd_3As_2$. Our work provides new physical insights into the intriguing transport properties of Dirac semimetals, revealing the importance of carrier density in mediating the linear MR in perpendicular fields and the quadratic negative MR in parallel fields. It also calls for the thin film growth of Dirac semimetals. By applying an external gate to effectively tune the Fermi level of the film toward the Weyl nodes, much larger linear MR and prominent negative MR would be expected.

## Methods

The $Cd_3As_2$ microribbon was grown by a chemical vapor deposition method. $Cd_3As_2$ powders and Si (001) covered with 2 nm Au layer were used as the precursor and substrates, respectively, and Argon were used as a carrier gas. Prior to each growth, the furnace was pumped and flushed for several times to remove water and oxygen using dry Argon. The precursor powder boat was placed in the hot center of the furnace, while the Si (100) substrates were placed down-stream about 32 cm away from the precursor powders. The furnace was gradually heated up to 750 ℃ in 20 min and the Ar flow was kept as 100 sccm during the growth process. The typical growth time is 60 min, and after then the furnace was cooled down to room temperature naturally. The high-resolution TEM identified [112] as the growth or axial direction of the ribbon. To study the magnetotransport properties, a four-terminal device was fabricated with standard e-beam lithography and lift-off processes. The transport properties of the device were then investigated in a Quantum Design PPMS system with the highest magnetic field up to 14 T.

## Acknowledgements

This work was supported in part by the Research Grants Council of the Hong Kong SAR under Grant Nos. 16305514, 17303714 and AoE/P-04/08, and in part by the National Natural Science Foundation of China under Grant Nos. 11204183 and 11374135. The electron-beam lithography facility is supported by the Raith-HKUST Nanotechnology Laboratory at MCPF (SEG HKUST08).

## Author contributions

 J.W. and S.S. conceived the project; H.L. grew the samples and fabricated devices with support from Z.F., H.Z., HC.L., and R.M.; H.H. and H.L. performed transport experiments; HZ.L. and S.S. provided theoretical support; H.H., HZ.L., H.L., J.W. and S.S. analyzed experimental data and wrote the manuscript with contribution from others.

## Additional information

**Competing financial interests:** The authors declare no competing financial interests.

## Figure Captions

**Figure 1. Nontrivial band structure and Berry curvature of a topological semimetal.** (a) A schematic of the energy spectrum of a topological semimetal. ($k_x$, $k_y$, $k_z$) is the wave vector. $k_{||}^2 = k_x^2 + k_y^2$ . (b) The vector plot of the Berry curvature in momentum space.

**Figure 2. Topological semimetal $Cd_3As_2$ microribbon device and magneto-transport characteristics.** (a) The SEM and (b) AFM images of the device. (c) The measured resistance ($R$) as a function of temperature ($T$) at zero magnetic field. (d) The magnetoresistance (MR) measured at 2 K with applied magnetic field ($B$) direction changing from perpendicular ($\theta = 90^0$) to parallel ($\theta = 0$) to the electric field ($E$) direction in the z-x plane. (e) The replot of MR with $\theta < 10°$ showing the negative MR at low magnetic fields.

**Figure 3. Linear MR in $B$ perpendicular to $E$.** (a) The MR measured at different $T$ indicated. (b) The $T$ dependence of the carrier mobility, which is calculated using Kohler's rule or Eq. (1). Inset: The MR at small magnetic fields and the parabola fittings (solid lines) at different $T$ indicated. (c) The carrier density as a function of $T$. The solid red curve is the fitting using Arrhenius's law. (d) The slope $k$ of the linear MR at high $B$ (from 8 to 14 T, obtained in (a) see solid curves) as a function of carrier density, $n$. The solid red and blue curves are the fittings using $n^{-1}$ and $n^{-2}$, respectively. The corresponding temperature for each data point is also indicated.

**Figure 4. Negative MR in $B$ parallel to $E$.** (a) The MR measured at different $T$ indicated. (b) The weak anti-localization effect at low temperatures and very small magnetic fields. (c) The positive conductivity $\Delta\sigma = \sigma - \sigma_0$, where $\sigma_0$ is the conductivity at $B = 0$, converted from measured negative MR at different $T$ indicated. The solid red curves are the fittings using $\sigma = C_a B^2$. (d) $C_a$ as a function of the carrier density, $n$. The solid red curve is a fitting using $n^{-2/3}$. The corresponding temperature for each data point is also indicated.

**Figure 5. MR in the quantum limit.** (a) $B_c$ as a function of $n^{2/3}$, $n$ is the carrier density. The corresponding temperature for each data point is also indicated. The solid

red line is a linear fitting. $B_c$ is defined as the critical field at which the MR in $B$ parallel to $E$ is minimum. At $B_c$, the system enters the quantum limit with a structure of the Landau bands illustrated as in the inset. $\nu$ is the index of the Landau bands and $\omega$ is cyclotron frequency. (b) The high $B$-field magnetoconductivity in in $B$ parallel to $E$ as a function of $1/B$. The magnetoconductivity is found to follow a $k'/B$ dependence while approaching 14 T (see solid red lines). (c) The slope $k'$ as a function of the carrier density. The solid red line is a linear fitting.

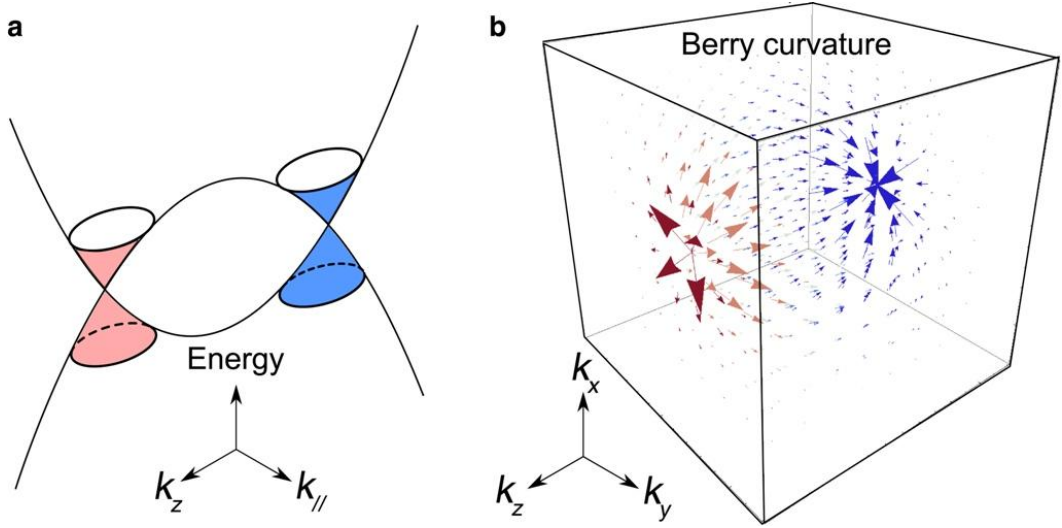

**Figure 1 Wang**

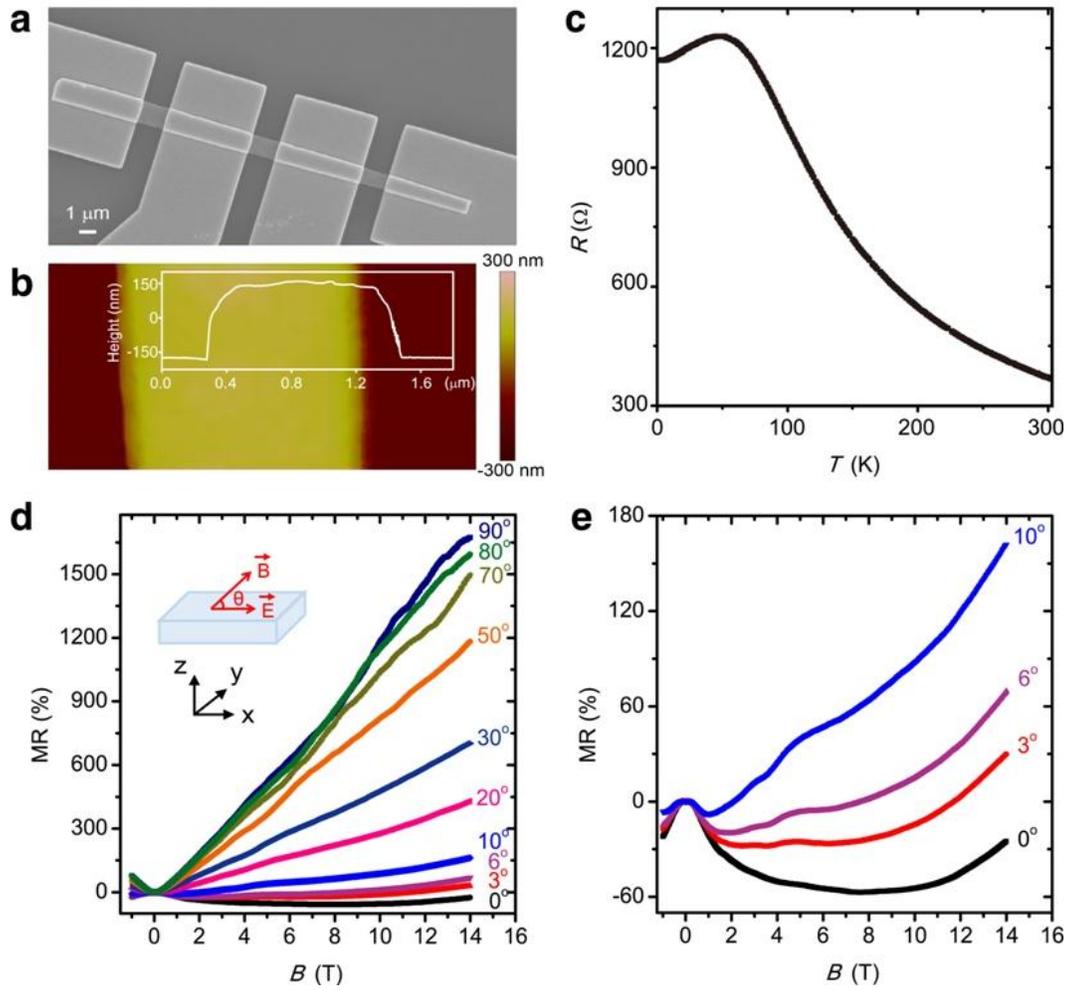

**Figure 2 Wang**

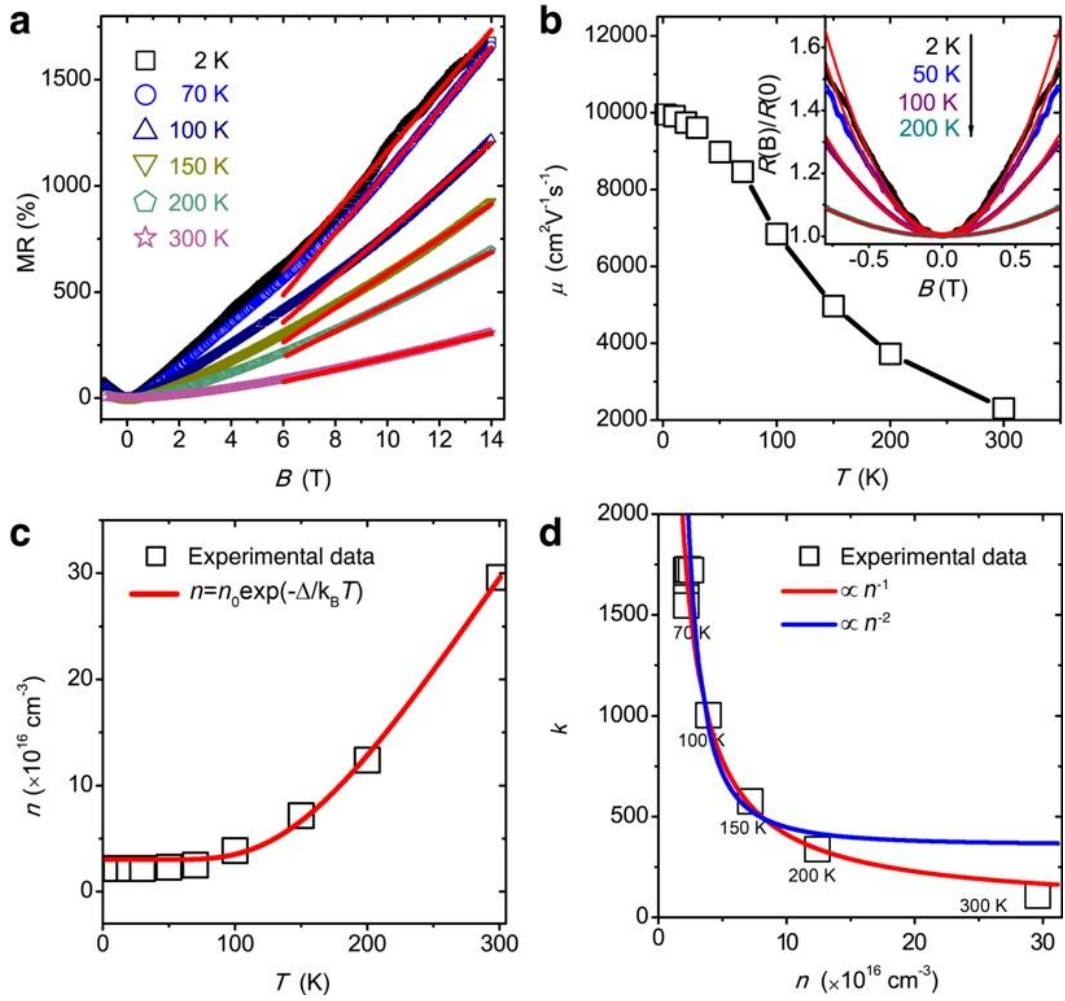

**Figure 3 Wang**

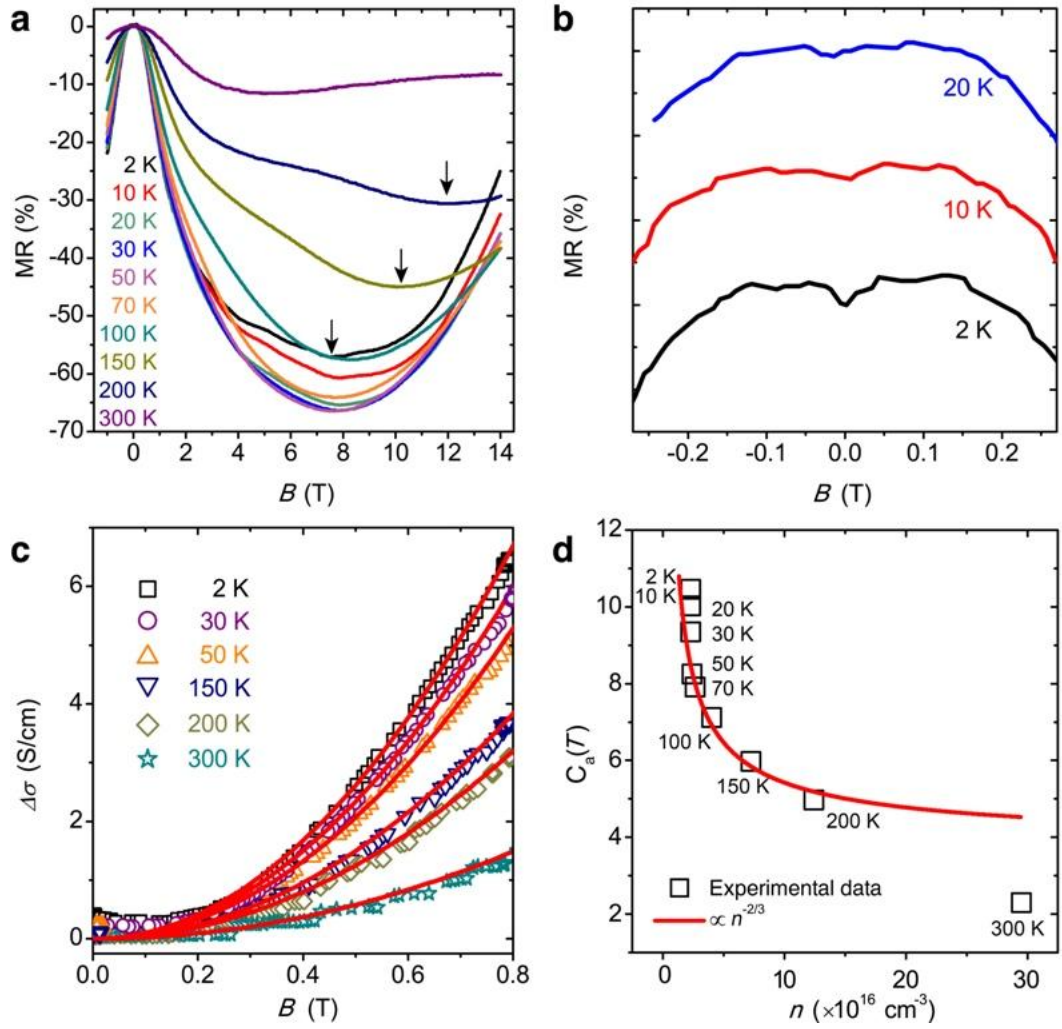

**Figure 4 Wang**

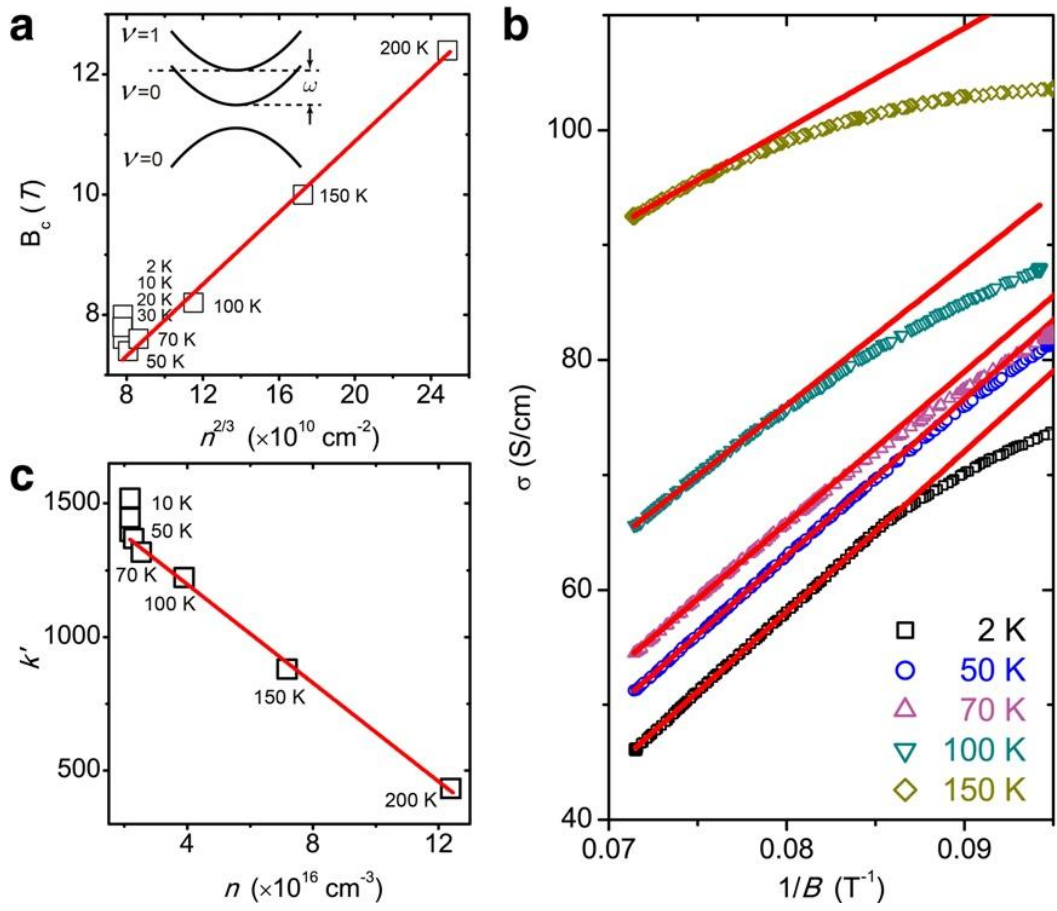

**Figure 5 Wang**